\documentclass[conference]{IEEEtran}
\IEEEoverridecommandlockouts
\usepackage{cite}
\usepackage{amsmath,amssymb,amsfonts}

\usepackage{graphicx}
\usepackage{textcomp}
\usepackage{xcolor}
\usepackage{algorithm}
\usepackage{algcompatible}
\usepackage{algpseudocode}

\usepackage{float}                  
\usepackage{subfig}                 
\usepackage{overpic}  

\def\BibTeX{{\rm B\kern-.05em{\sc i\kern-.025em b}\kern-.08em
    T\kern-.1667em\lower.7ex\hbox{E}\kern-.125emX}}

\makeatletter
\newcommand{\linebreakand}{%
  \end{@IEEEauthorhalign}
  \hfill\mbox{}\par
  \mbox{}\hfill\begin{@IEEEauthorhalign}
}
\makeatother
\begin{document}

\title{1-D CNN-Based Online Signature Verification with Federated Learning\\
{\footnotesize \textsuperscript{}}
\thanks{This research was partially supported by KAKENHI (Grant-in-Aid for JSPS Fellows) 22KJ0614.}
}

\author{\IEEEauthorblockN{1\textsuperscript{st} Lingfeng Zhang}
\IEEEauthorblockA{
\textit{The University of Tokyo}\\
Tokyo, Japan \\
zhang-lingfeng936@g.ecc.u-tokyo.ac.jp}
\and
\IEEEauthorblockN{2\textsuperscript{nd} Yuheng Guo}
\IEEEauthorblockA{
\textit{The University of Tokyo}\\
Tokyo, Japan \\
yuhengguo@satolab.itc.u-tokyo.ac.jp}
\and
\IEEEauthorblockN{3\textsuperscript{rd} Yepeng Ding}
\IEEEauthorblockA{
\textit{The University of Tokyo}\\
Tokyo, Japan \\
youhoutei@satolab.itc.u-tokyo.ac.jp}
\linebreakand
\IEEEauthorblockN{4\textsuperscript{th} Hiroyuki Sato}
\IEEEauthorblockA{
\textit{The University of Tokyo}\\
Tokyo, Japan \\
schuko@satolab.itc.u-tokyo.ac.jp}
}

\maketitle

\begin{abstract}
Online signature verification plays a pivotal role in security infrastructures. However, conventional online signature verification models pose significant risks to data privacy, especially during training processes. To mitigate these concerns, we propose a novel federated learning framework that leverages 1-D Convolutional Neural Networks (CNN) for online signature verification. Furthermore, our experiments demonstrate the effectiveness of our framework regarding 1-D CNN and federated learning. Particularly, the experiment results highlight that our framework 1) minimizes local computational resources; 2) enhances transfer effects with substantial initialization data; 3) presents remarkable scalability. The centralized 1-D CNN model achieves an Equal Error Rate (EER) of 3.33\% and an accuracy of 96.25\%. Meanwhile, configurations with 2, 5, and 10 agents yield EERs of 5.42\%, 5.83\%, and 5.63\%, along with accuracies of 95.21\%, 94.17\%, and 94.06\%, respectively.
\end{abstract}


\section{Introduction}
A handwritten signature is a biometric method used for personal identification\cite{2023_survey}. Long in history, handwritten signatures have served as a means for individuals to endorse documents. Nowadays, handwritten signatures continue to be essential in modern forensic science, while digital signatures have made significant strides \cite{SSI}. As a behavioral attribute, a handwritten signature, along with voice, gait, and other behavioral traits, contributes to behavioral biometrics. Unlike physiological attributes such as fingerprints, palm prints, DNA, and facial features, handwritten signatures capture individual behavior patterns. These patterns exhibit low intra-person variance, as an individual's signatures tend to be consistent or similar over time. However, they also show high inter-person variance, meaning that signatures of different individuals possess noticeable differences. Behavioral and physiological attributes together form the basis of biometric identification for individuals. We attribute the widespread application of handwritten signatures to their high accuracy and ease of use. The high accuracy stems from the uniqueness of a handwritten signature to an individual and the difficulty of being transferred, lost, or stolen \cite{2023_survey}. Therefore, handwritten signatures are an effective biometric means for personal identity recognition.

Regarding the data modality, we classify signatures into two categories: offline and online. Unlike offline signatures \cite{offline_1} \cite{offline_2}, which only consider the static image of the signature, online signatures \cite{online_1}\cite{online_2}\cite{online_3}\cite{online_4}\cite{2023_survey} include dynamic features such as, x- and y-coordinates, pen pressure, pen angle, pen rotation angle, writing speed, etc. Online signatures are a collection that encapsulates the signature itself and the dynamic historical records while writing the signature. Due to the increased complexity of online signatures and the intricate dependencies among different features, they are difficult to imitate, thus providing higher security.

The acquisition of data poses a significant challenge in online signature verification tasks. One key challenge is that each individual possesses a unique signature style, implying the need for a substantial number of signatures to capture the diversity and variations in signature styles. It is worth noting that the collected signature distribution may not necessarily represent the actual signature distribution accurately. Therefore, training an online signature verification system requires substantial data to approximate the real distribution effectively. However, signature data's sensitivity and privacy concerns restrict its shareability. Personal signatures are typically regarded as sensitive information, necessitating strict legal and ethical frameworks for safeguarding user privacy. Consequently, aggregating multiple databases into a centralized database for model training is a formidable task, despite the potential benefits a centralized database could offer in terms of enhancing model performance. 

To address this issue, specifically to harness a vast amount of data effectively while ensuring user privacy protection, we propose a framework using federated learning \cite{FL} for online signature verification. We incorporate a 1-D CNN into the federated learning framework. As shown in Figure \ref{fig:overview}, We propose a novel application scenario where, within the entire framework, a central coordinator is in the cloud, and several local agents are located locally. The local agents represent independent entities with access to a local signature dataset and a need for signature authentication. For instance, one entity could represent the local bank department responsible for managing check signatures, or one could represent the front of a local supermarket responsible for managing credit card signatures. Rather than aggregating all the datasets of agents for training a central model, federated learning utilizes a central coordinator to manage the configurations and updates of a global model. Each agent conducts local training and uploads its model's weights to the coordinator for aggregation during an iteration. It enables the global model to adapt to the various local database distributions. Additionally, the choice of 1-D CNN ensures that the model can extract effective features, and its lightweight weight parameters facilitate communication and training.


Our contributions are summarized as follows:
\begin{itemize}
\item We propose a federated learning framework to address privacy challenges in online signature verification.
\item We integrate a 1-D CNN model in our framework to reduce the overall computation cost by a lightweight algorithm.
\item We demonstrate the effectiveness of our 1-D CNN model and federated learning approach through extensive experiments regarding lightweight local training, improved transfer effect, and high scalability.
\end{itemize}

\section{Related Work}
A signature is composed of symbols and strokes that represent the behavioral characteristics of the writer. Based on appropriate feature extraction, the feature space can reflect whether a signature from a specific user is genuine or forged. Online signature verification differs from offline signature authentication because it deals with dynamic features, including writing speed, angles, pen pressure, and more. As a result, it possesses higher accuracy and security. 

\textit{Online signature verification} is a process to ascertain the authenticity of a signature by analyzing dynamic features recorded during the signing procedure. This process encompasses various methodological approaches to attain precise and secure outcomes.

In traditional contexts, two frequently utilized methodologies are Dynamic Time Warping (DTW \cite{DTW}) \cite{DTW3}\cite{DTW4}\cite{DTW1} and Hidden Markov Models (HMM \cite{DTW}) \cite{HMM2}\cite{HMM3}. DTW stands out for its capacity to robustly compare signature dynamics while accommodating temporal variations. In contrast HMMs are well-suited for modeling sequential data because they capture the dependencies between consecutive data points.
Online signature data often exhibits intricate dependencies, where current observations are contingent on past observations and data from other channels. Deep learning, renowned for capturing non-linear features, has gained substantial prominence in recent years in online signature verification. For instance, in the studies by \cite{RNN1}\cite{RNN2}\cite{RNN3}, Recurrent Neural Networks (RNNs) are employed due to their effectiveness in capturing time-series features and long-term dependencies, thus facilitating the recognition of dynamic signature patterns and individualized characteristics.

The distinction between our work and the aforementioned methods lies in utilizing a simple 1-D Convolutional Neural Network (CNN) architecture for handling temporal tasks. Subsequent experiments will demonstrate that larger kernel sizes can enhance the 1-D CNN model's capacity to process temporal data. Another distinguishing difference is that we do not necessitate a centralized database because we adopted the federated learning \cite{FL} framework.

\section{Proposed Framework}
\subsection{Overview}
In general, we present a novel lightweight online signature verification framework built on the principles of federated learning \cite{FL}. Specifically, we embed an 1-D convolutional neural network into the federated learning framework, enabling multiple agents to jointly train a global model that adapts to their respective local data distributions without sharing privacy-sensitive signature data. Federated learning allows us to keep privacy-sensitive data decentralized among the clients. Each client performs local independent training on its local dataset. The desired global model evolves in each iteration by aggregating information reflecting local datasets; in our case, the communication involves the local models' weights rather than the raw data of the datasets. This approach provides the advantage of solving the original online signature verification goal while preserving the privacy of the signature users, as their sensitive data never leave the local environment. The main benefit of federated learning is the significant reduction of privacy and security risks by confining the hazards to the local level rather than centralizing all data in a single data center \cite{FL}.

\begin{figure*}[htbp]
\centerline{\includegraphics[width=0.6\textwidth]{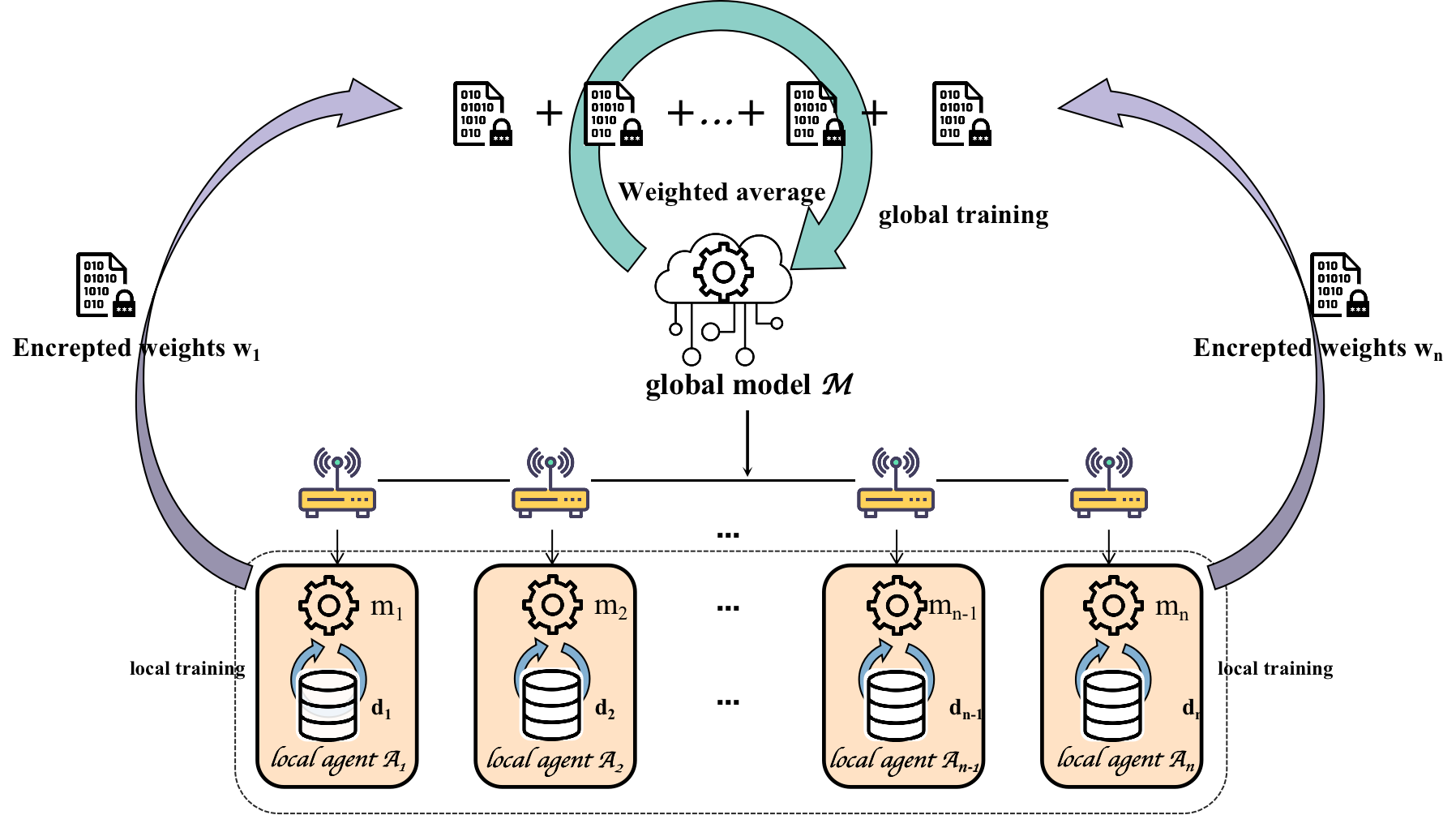}}
\caption{Overview of our proposed framework.}
\label{fig:overview}
\end{figure*}

Figure \ref{fig:overview} overviews our framework. There are $n$ local agents $\mathcal{A}_i \ (i = 1,2,...,n)$ at the edge, and each has its own local dataset, denoted as $d_i$. Each client simulates a real-world location where signature verification, such as a local authentication center, needs to be performed. 

Following \cite{FL}, we adopt a synchronous strategy in each iteration to update the parameters \textemdash All local agents perform local training in each iteration. In each iteration, the local agent fetches the latest model parameters from the central coordinator on the cloud. The agent then conducts independent local training, including data preprocessing and backpropagation. After $e$ epochs of training, the central coordinator receives the model parameters from each local agent through communication and performs aggregation to update the global model's parameters. The entire process of fetching local data, performing local training, communicating with the central coordinator, aggregating the model updates, and updating the global model repeats $E$ times, where $E$ denotes the number of iterations.

\begin{figure}[htbp]
\centerline{\includegraphics[width=0.18\textwidth]{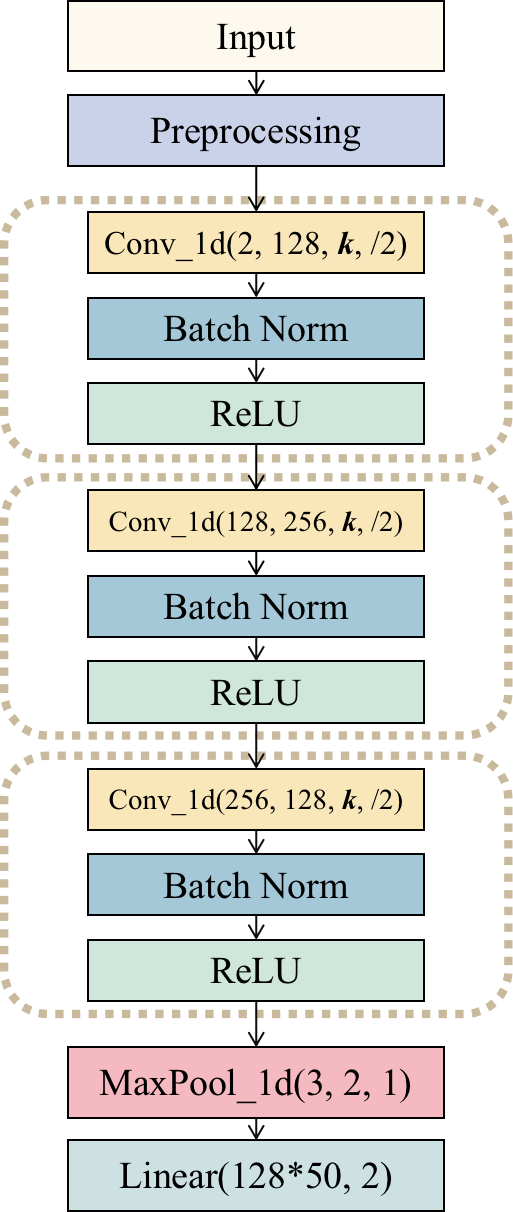}}
\caption{Model configurations.}
\label{fig:config}
\end{figure}

\subsection{Verification Model Design}\label{model_config}
The global model is the central part of our framework, shown at the center in Figure \ref{fig:overview}. The global model serves as our primary verification model and is initialized in the coordinator in the cloud. Every agent downloads the global model, performs local training, and then uploads the updated model weights in each iteration. Finally, the coordinator executes some aggregation algorithms, e.g., FederatedAveraging. It means that the global and downloaded local models are homogenous at the beginning of each iteration.

One input signature $ \mathbf{X}_t^i = ( \mathbf{x}_1,  \mathbf{x}_2, ...,  \mathbf{x}_{T_i})$ is a vector in space $\mathbb{R}^{c \times T_i}$, where the index $t$ represents the temporal dependency, $i$ represents the index of the signature, thus $T_i$ represents the length of signature with index $i$. $\mathbf{x}_t \ (t = 1, 2, ..., T_i)$ is a $c$-dimensional vector, each element of which represents a specific observation of a channel at timestamp $t$. For example, consider a digital pencil equipped with 4 sensors. The first recorded Signature $\mathbf{X}_t^1$ has a length of $T_1$. The observation at time point $t = 3$, denoted as $\mathbf{x}_3$, is a 4-dimensional vector representing the recorded data from these 4 sensors at $t = 3$. It is worth noting that different signatures generally possess varying lengths.
\par
The verification model aims to produce a predicted label $\mathbf{Y}^i$ based on the input $\mathbf{X}^{i}_{t}$, where $\mathbf{Y}^i$ $\in$ $\{Genuine, Forged\}$. Figure \ref{fig:config} illustrates the model configurations used for verification. The input undergoes a preprocessing stage, where it is normalized and padded to a fixed length for application to the CNN layers. Subsequently, the features pass through three CNN blocks. Each CNN block consists of a 1-D CNN layer, a batch normalization layer, and a ReLU activation function layer. Additionally, the feature's length is halved from its original size after each CNN block. In practice, we used a dataset with a maximum length of 713, so we zero-padded it to a length of 800. After passing through the entire 1-D CNN model, the length is reduced to 50. Furthermore, the downsampled features are passed through a 1-D MaxPool layer. Finally, the concatenated feature maps are processed by a linear layer to output the labels.

\subsubsection{\textbf{Preprocessing}}
In the preprocessing stage, we have two main steps: normalization and zero-padding. The preprocessing corresponds to the purple box in the diagram in Figure \ref{fig:config}.

Firstly, we perform normalization on an input signature. We adopt common normalization \cite{normalize}. It refers to the process of transforming signature data into a standardized format before further feature extraction. In our case, we only process the x and y coordinates of the signature, denoted as $\mathbf{X}_t^{i} = ( \mathbf{x}_1,  \mathbf{x}_2, ...,  \mathbf{x}_{T_i}) = ((x_1, y_1), (x_2, y_2), ..., (x_{T_i}, y_{T_i}))$, where $(x_t, y_t)$ represents the x- and y-coordinate values at time $t \ (t = 1, 2, ..., T_i)$.
\begin{equation}
    x_t \leftarrow \frac{x_t - x_g}{x_{max} - x_{min}},
    y_t \leftarrow \frac{y_t - y_g}{y_{max} - y_{min}},
\end{equation}
where $(x_g, y_g)$ represents the centroid of the signature, and $(x_{min}, y_{min})$ and $(x_{max}, y_{max})$ are the minimum and maximum of $(x_t, y_t)$ for $i = 1, 2, ..., T_i$. Normalization reduces the impact of variations in writing styles, sizes, and positions of signatures. It enhances the system's ability to handle different signature variations and increases its robustness across diverse signatures.

Secondly, CNNs commonly use mini-batches \cite{minibatch} for training. To facilitate batch processing, sequences within a batch must have the same length. By adding zeros at the end of a sequence\textemdash zero padding \textemdash we ensure that all input sequences to a CNN layer have the same size, enabling efficient parallel processing. i.e., 
\begin{equation}
    \mathbf{X}_t \leftarrow (\mathbf{X}_t, \underbrace{0,0,...,0}_{T_{max} - T_i}),
\end{equation}
where $\mathbf{X}_t$ represents the normalized x- and y-coordinates as detailed previously, and $T_{max}$ indicates the largest length across all signatures.

\subsubsection{\textbf{Convolutional Blocks}}
Since its introduction, the 1-D Convolutional Neural Networks (CNNs) have shown the state-of-the-art performance in time-series-related tasks \cite{1d_CNN_survey}, such as real-time electrocardiogram (ECG) monitoring \cite{EEG1}\cite{EEG2}. The significant advantage is their low-cost implementation. It is due to the simple and compact configuration of 1D CNNs, which perform only 1D convolutions involving scalar multiplications and additions. The advantages of 1-D CNN lie in its ease of training \cite{1d_CNN_survey}, which implies fewer communication rounds and a smaller number of parameters, resulting in reduced communication overhead. Our study uses a simple 1-D CNN to verify online signatures. Despite its simplicity, the model exhibited impressive results in accomplishing the task.

The convolutional blocks correspond to the part enclosed by the dashed line in the diagram in Figure \ref{fig:config}. Each convolutional block consists of a 1D convolutional layer, a batch normalization layer, and a ReLU activation function.
\par
The \textbf{1-D CNN layer} creates feature maps entailing learnable filters (kernels) that convolve over the input sequence, aiming to detect distinct patterns or features. Each filter undergoes element-wise multiplication with the input's local region, culminating in a single value within the output feature map, signifying the presence of the identified feature. In place of perceiving the layer as a singular vector-to-vector function, an alternative perspective arises, wherein the layer comprises neural units functioning in parallel. Each unit assumes the role of representing a vector-to-scalar function \cite{DL}:
\begin{equation}
    \mathbf{s}_{j}^{l} \leftarrow \mathbf{b}_{j}^{l}+\sum_{i=1}^{k}conv1D(\mathbf{w}_{ij}^{l-1}, \mathbf{s}_{i}^{l-1}),
\end{equation}
where $\mathbf{s}_{j}^{l}$ represents the value of the $j$-th neuron in layer $l$ for the 1-D CNN layer, $\mathbf{b}_{j}^{l}$ represents the bias of $j$-th neuron in layer $l$, $\mathbf{w}_{ij}^{l-1}$ is the kernel weight from $i$-th neuron in layer $l-1$ to $j$-th neuron in layer $l$, $k$ is the size of kernel, and $conv1D$ is the convolution operation without zero padding.
\par
We employ \textbf{batch normalization} \cite{BN} to enhance training stability and improve the convergence speed during the training process. The batch normalization principle involves normalizing a layer's input by fixing its mean and variance through a normalization operation.
\begin{equation}
    \mathbf{s}_{q}^{l} \leftarrow \frac{\mathbf{s}_{q}^{l} - \mathbf{\mu}_{\mathcal{B}}}{\sqrt{(\mathbf{\sigma}_{\mathcal{B}})^2 +\epsilon}},
\end{equation}
\begin{equation}
    \mathbf{s}_{q}^{l} \leftarrow \mathbf{\gamma}\mathbf{s}_{q} + \mathbf{\beta},
\end{equation}
where $\mathcal{B}$ represents a mini-batch with $n_{\mathcal{B}}$ samples, $q \ (q = 1,2,...,n_{\mathcal{B}})$ is the index for neuron values $\mathbf{s}^l$ at layer $l$ across the mini-batch, and $\gamma$ and $\beta$ are learnable parameters for the batch normalization layer.
\par
The activation function is the \textbf{Rectified Linear Unit (ReLU)} at the end of the CNN block, i.e. for neuron $\mathbf{s}_j^{l}$,
\begin{equation}
    f(\mathbf{s}_j^{l}) = max(0, \mathbf{s}_j^{l}) = \frac{\mathbf{s}_j^{l} + |\mathbf{s}_j^{l}|}{2}.
\end{equation}
ReLU is computationally efficient and helps alleviate gradient vanishing, leading to better gradient propagation \cite{ReLU}.
\subsubsection{\textbf{MaxPool Layer}}
The 1-D MaxPool layer corresponds to the red box in Figure \ref{fig:config}. We employ a 1-D MaxPool layer for downsampling, which reduces the dimensionality of the features. 
\begin{equation}
    \mathbf{s}^{l} \leftarrow \mathbf{s}^{l} \downarrow ss,
\end{equation}
where $ss$ denotes the downsampling operation with a scalar factor $ss$. In our case, the window size for the 1-D MaxPool layer is 3 with a stride of 2. This approach provides several advantages, as it captures the maximum value within a specified window, promoting translational invariance to small changes in the input. As a result, the network becomes more robust and capable of generalizing better to variations in the data.

\subsubsection{\textbf{Linear Layer}}
Finally, our model concludes with a linear layer that maps the feature maps to a 2-dimensional output representing the labels. A fully connected or linear layer is mainly for performing linear transformations on the feature map. The linear layer maps the feature map to an output space, where each output neuron is connected to every input neuron through weighted connections. It allows the layer to learn and represent complex patterns and relationships in the data.
\begin{equation}
    \mathbf{y} \leftarrow \mathbf{W} \mathbf{s} + \mathbf{b},
\end{equation}
where $W \in \mathbb{R}^{128 \cdot 50 \times2}$ and $b \in \mathbb{R}^{m \times2}$ are the parameters for the linear layer, and $n_{\mathcal{B}}$ is the size of mini-batch. 128 represents the final number of output channels in Figure \ref{fig:config}, 50 is the final length of inputs. 

\subsection{Federated Learning-Based Model Training}
\subsubsection{\textbf{Local Training}} \label{local_train} 
We refer to independent local training as the process in federated learning where participating agents train the model locally, and only the model's updated parameters are returned to the central server.
\par
We set the loss function to be the cross-entropy. The cross-entropy loss is commonly used in classification tasks and measures the dissimilarity between the predicted probability distribution and the true label distribution. 
\begin{equation}
    L(\mathbf{X}_t, \mathbf{Y})= - (y\cdot log(p)+(1-y) \cdot log(1-p)),
\end{equation}
where y is the label, which is either $0$ or $1$ in our case, $p$ is the predicted probability of the \textit{Genuine}.
\par
We utilize the Stochastic Gradient Descent (SGD) algorithm \cite{SGD}\cite{SGD2} to update the model parameters during training. It aims to minimize the loss function by iteratively updating the model's parameters in the direction of the steepest descent of the loss surface. 
\begin{equation}
    \mathbf{w} \leftarrow \mathbf{w} - \eta \nabla_{\mathbf{w}} L(\mathbf{X_t}, \mathbf{Y}, \mathbf{w}) = \mathbf{w} - \frac{\eta}{n_{\mathbf{B}}}\sum^{n}\nabla L_{\mathbf{w}}(\mathbf{X_t}, \mathbf{Y}, \mathbf{w}),
\end{equation}
where $\eta$ is the learning rate, $\nabla_{\mathbf{w}} L$ is the gradients computed by backpropagation algorithm \cite{DL}.

\subsubsection{\textbf{Federated Training}}
We derive our federated training framework by building upon the \textit{FederatedAveraging} algorithm \cite{FL}. \textit{FederatedAveraging} can be seen as a variant of the SGD algorithm, focusing on training deep neural networks through iterative model averaging. Compared to centralized training, \textit{FederatedAveraging} decouples the model's training from direct access to the raw data, providing privacy advantages. Simultaneously, compared to the naive application of SGD in the context of \textit{FederatedSGD}, where gradients are communicated, \textit{FederatedAveraging} significantly reduces communication volume at the expense of some local computation cost \cite{FL}.

Algorithm \ref{alg:fedavg} summarizes our federated training framework. The coordinator located at the central server initializes a global model $w_0$. In each iteration, the agents at the edge parallelly conduct independent training based on their local databases and then communicate the model parameters to the central server. The central server computes a weighted average of the new model parameters to obtain a new global model, where the weights are proportional to the sizes of the agents' databases. Each agent trains the model according to the discussed local training process in section \ref{local_train}. The iteration repeats.

\begin{algorithm}
\caption{\textit{Federated training framework. }.}\label{alg:fedavg}
\begin{algorithmic}
\Require $w_i$ indicates model at iteration $i$. $k$ is the index for agents. $E$ indicates the number of local epoches. $p_k$ denotes the training set for agent $k$.\\
\textbf{Server executes:} \\
   \hskip1.0em initialize $w_0$ \\
    \hskip1.0em\textbf{For} {iteration $i$ = 1,2, ..., $I$} \textbf{do}\\
    \hskip2.0em\textbf{For} {agent $k$ = 1,2, ..., $K$} \textbf{do}\\
    \hskip3.0em $w_{i+1}^{k} \leftarrow $ Local\_Training($k, w_t$)\\
    \hskip2.0em $w_{i+1} \leftarrow \sum_{k}\frac{\mathcal{P}_k}{\mathcal{P}}w_{t+1}^{k}$\\
    \\
\textbf{Local\_Training} // run on client $k$\\
\hskip1.0em $\mathcal{B} \leftarrow$ split $\mathcal{P}_k$ into batches of size $\mathcal{B}$\\
\hskip1.0em \textbf{For} {each local epoch i = 1,2, ..., E} \textbf{do}\\
\hskip2.0em \textbf{For} {batch $b \in B$} \textbf{do}\\
\hskip3.0em $w \leftarrow \eta \nabla L(\theta, w)$ // equation (11) \\
\hskip1.0em return $w$ to the server
\end{algorithmic}
\end{algorithm}

\section{Experiments}

\subsection{Datasets \& Platform}
We validate our framework on the public SVC-2004\cite{SVC-2004} online signature dataset. Two tasks compose the dataset, namely Task 1 and Task 2. While Task 1 only records coordinate information, Task 2 encompasses additional information such as pen orientation and pressure besides coordinate information. Each task consists of 40 users, with 40 signature samples representing each user. Among these samples, 20 are genuine signatures, while the remaining 20 are forged signatures from skilled forgeries. To clarify, within each task, there contribute 1600 signatures contributed by 40 users. Considering the entirety of the project, there exist 3200 signatures in total, originating from a collective of 80 users. SVC-2004 is a benchmark dataset in the online signature verification domain. In our experiments, we only consider utilizing the x-y coordinates.

The experimental platform employed in our experiments comprises a hardware configuration, including AMD Ryzen 9 7950X 16-Core Processor and The NVIDIA GeForce RTX 4090 graphic card. We train and analyze models on Python 3.10.11, Pytorch 2.0.0, and Ubuntu 22.04.2. We choose ROC curves \cite{ROC} as our primary metric. Receiver Operating Characteristic curves (ROC curves) visually illustrate the trade-off between a classification model's true positive rate and false positive rate, offering a graphical representation of its performance across different discrimination thresholds. 

The core components we aim to validate in our experiments are the verification model based on 1-D CNN and the federated learning framework based on Federated Averaging. In the section \ref{1-D CNN} of our experiments, we validate the verification model and explore the impact of different \textit{kernel sizes}. 
In the section \ref{fed_vali}, we validate our proposed online signature verification framework based on federated learning.

\subsection{Model Effectiveness} \label{1-D CNN}

We validate the effectiveness of the 1-D CNN as a verification model in a centralized training manner. We examine the sensitivity of our verification model under different \textit{kernel sizes}.

\begin{figure}[h]
\centerline{\includegraphics[width=0.40\textwidth]{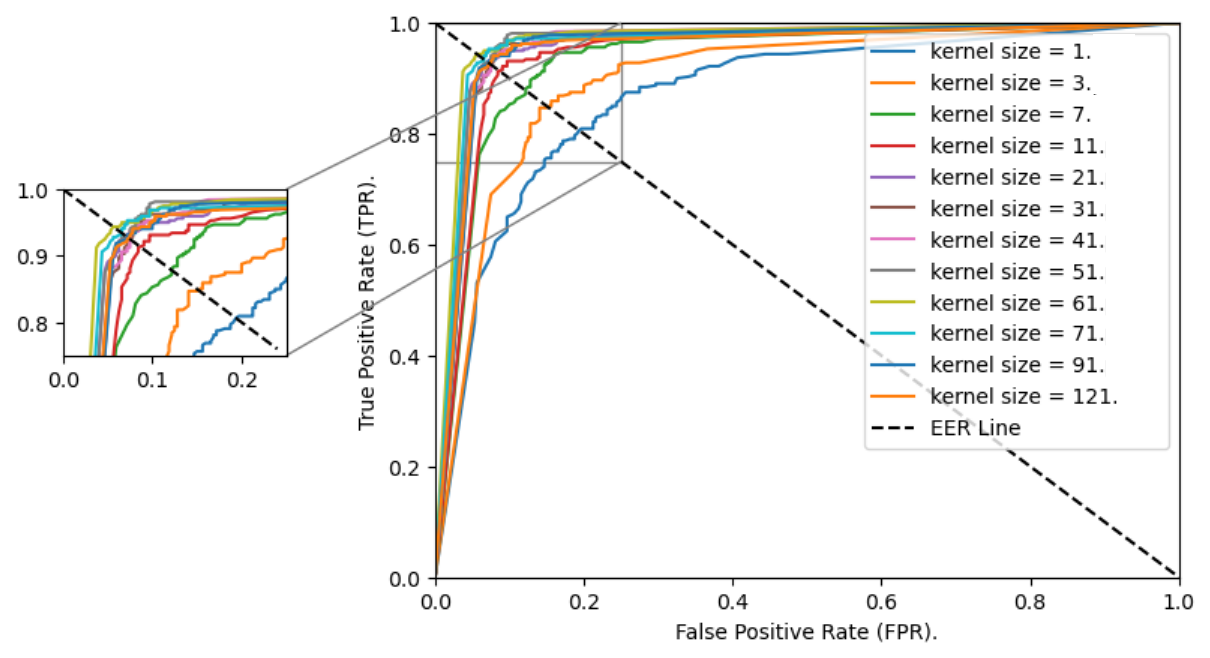}}
\caption{EER for different kernel sizes.}
\label{fig:centr_EER}
\end{figure}

We combined Task 1 and Task 2 of the SVC-2004 database, resulting in a total of 80 users, with each user providing 20 genuine signatures and 20 forged signatures. It amounts to a total of 3200 signatures. We randomly sampled 16 genuine and 16 forged signatures for each user as training data. The remaining 4 genuine and 4 forged signatures are reserved as a test dataset for model validation. The ratio of our training dataset to the test dataset is 8:2. We use the Adamax optimizer\cite{adam} with an initial learning rate of 0.01 to train our model. Adamax is a variant of the Adam optimizer that incorporates self-adaptive gradients, allowing for efficient and effective model training. Figure \ref{fig:centr_EER} presents the EER for each model. It is evident that with an increasing kernel size, both accuracy and EER show gradual improvement until reaching a peak at size 51 for accuracy and 61 for EER. Subsequently, we fix 61 as the \textit{kernel size} for our verification model.

\begin{figure}[h]
\centerline{\includegraphics[width=0.35\textwidth]{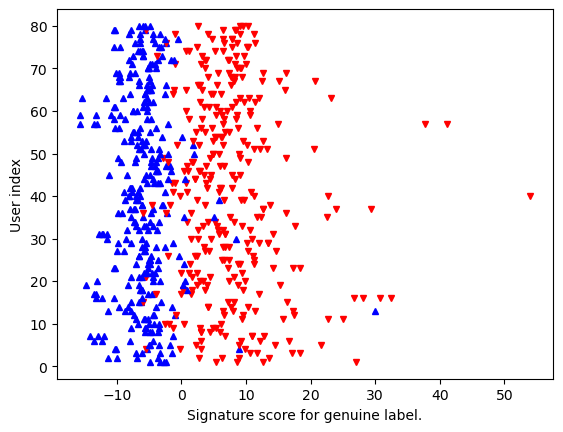}}
\caption{Test scores in one experiment where \textit{Kernel size = 61}.}
\label{fig:centr_score}
\end{figure}

We set the number of training epochs to 200 and \textit{batch size} to 160. Figure \ref{fig:centr_EER} presents the EER for each model. It is evident that with an increasing kernel size, both accuracy and EER show gradual improvement until reaching a peak at size 51 for accuracy and 61 for EER. Subsequently, we fix 61 as the \textit{kernel size} for our verification model. Figure \ref{fig:centr_score} presents the scores of the model with a kernel size of 61 on the test set for various signatures. Blue triangles denote genuine signatures, while red triangles represent skilled forgeries. Each row corresponds to a unique user in the SVC-2004 dataset.

\subsection{Federated Learning Effectiveness} \label{fed_vali}

We employ a heuristic parameter search method for validation. Heuristic parameter search refers to independently exploring an isolated parameter's impact while keeping other parameters constant. Here, we explored the effects of \textit{local epochs} $E$, \textit{initial dataset size}, and the \textit{number of agents} $K$ on the framework respectively.

\subsubsection{\textbf{Lightweight}}

\begin{figure}[h]
\centerline{\includegraphics[width=0.40\textwidth]{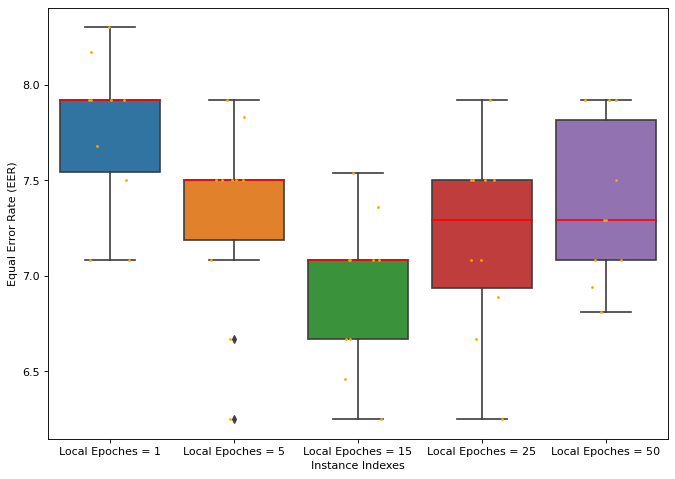}}
\caption{Boxplot for different \textit{local epochs}.}
\label{fig:local_epochs_box}
\end{figure}

\begin{figure}[h]
\centerline{\includegraphics[width=0.40\textwidth]{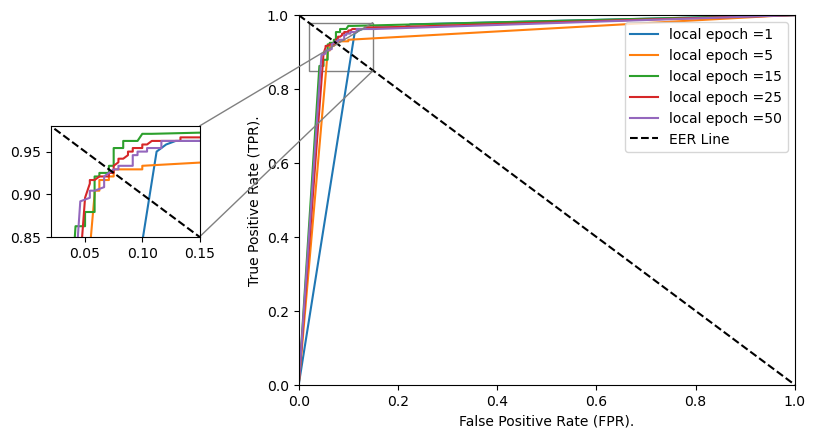}}
\caption{ROC curves for different \textit{local epochs}.}
\label{fig:local_epochs_EER}
\end{figure}

\begin{figure}[h]
\centerline{\includegraphics[width=0.40\textwidth]{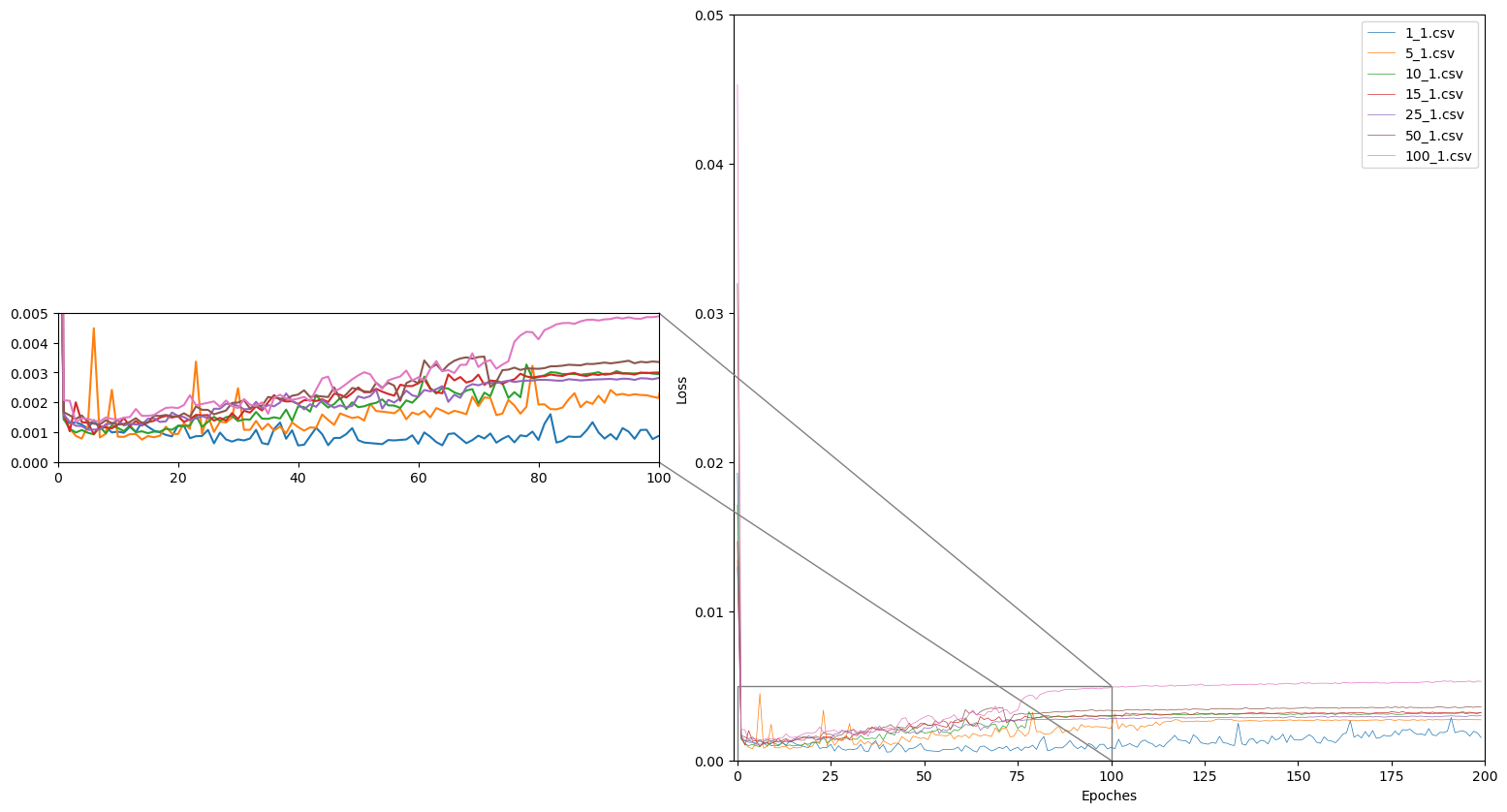}}
\caption{Training loss for different \textit{local epochs}.}
\label{fig:local_epochs_loss}
\end{figure}

We investigate the impact of different \textit{local epochs} $E$ on the model's performance. \textit{Local epochs} refer to the number of training epochs conducted on each agent locally. We set the dataset size for initialization $\mathcal{P}_{init}$ to 320, corresponding to $20$ users, each providing $16$ samples. We set $K$ to $2$, meaning the remaining 60 users are evenly divided into two groups, each containing 30 users. Each user contributes 16 samples for the training set and 4 samples for the validation set, resulting in $\mathcal{P}_k$ = 480, $k=1,2$. We set the number of \textit{iterations} $I$ to 200, the \textit{local batch size} $\mathcal{B}_{local}$ to 32, and the \textit{learning rate} $\eta$ to 0.001. The search space for the \textit{local epochs} $E$ is $\{1, 5, 15, 25, 50\}$. Figure \ref{fig:local_epochs_box} compares different parameter values, while Figure \ref{fig:local_epochs_EER} presents the ROC curves corresponding to each parameter value. Please note that for each parameter in Figure 1, we have conducted 10 model instances, and the ROC curves plotted in Figure 2 represent the median instance. We can observe that the model performs optimally on the validation set when \textit{local epochs} $E = 15$. It is worth mentioning that a larger value of local epochs does not necessarily guarantee better model performance. An empirical explanation for this phenomenon is that excessive local epochs may lead to overfitting the model on local data, thus reducing the model's generalization performance. Figure \ref{fig:local_epochs_loss} illustrates the loss values for different \textit{local epochs} along training epochs. We can observe that the \textit{local epochs} have a limited impact on the convergence of the model.

Combining Figures \ref{fig:local_epochs_box} and \ref{fig:local_epochs_EER}, we observe that only 15 local epochs are needed for local training. Furthermore, the qualitative analysis in Figure 9 reveals that varying the number of local epochs does not increase the global iteration rouns. \textbf{It implies that we can perform local model training with relatively low local computational requirements without significantly increasing the number of global iterations, thus highlighting the lightweight nature of our model}.

\subsubsection{\textbf{Transfer Effect}}
We can perceive federated learning as a form of transfer learning, transferring knowledge from the source domain \textemdash the signature distribution provided by volunteers at the center \textemdash to the target domain \textemdash the distribution of signatures managed by decentralized local agents. Hence, the training data of the initial model holds a pivotal role, influencing factors such as the model's overall performance, generalization capability, and transfer effects.

\begin{figure}[h]
\centerline{\includegraphics[width=0.40\textwidth]{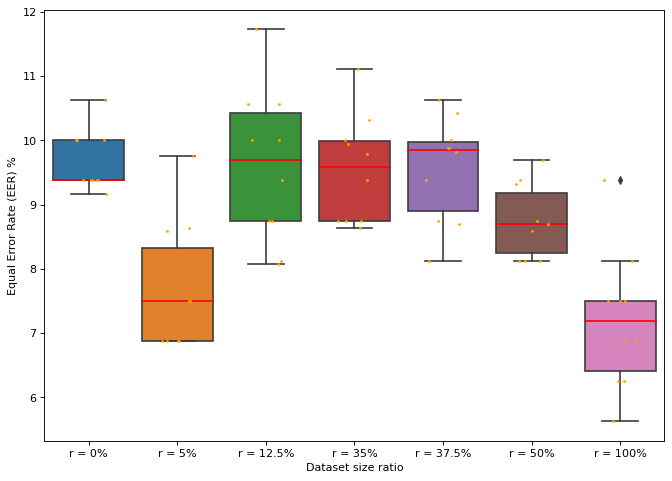}}
\caption{Boxplot for different \textit{initial dataset size}.}
\label{fig:initial_box}
\end{figure}

\begin{figure}[h]
\centerline{\includegraphics[width=0.40\textwidth]{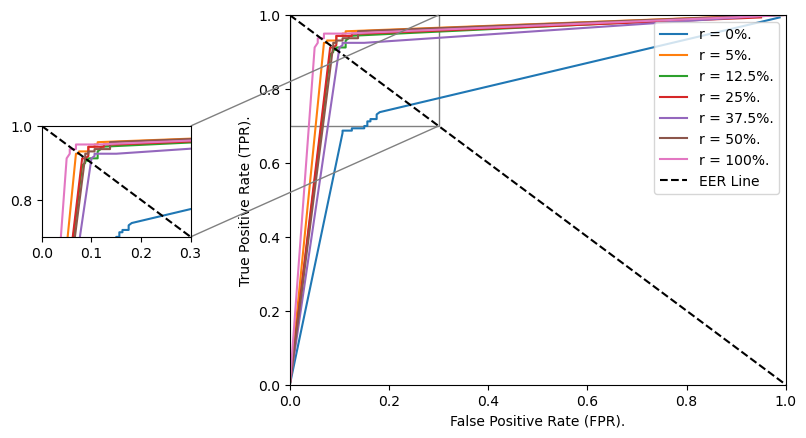}}
\caption{ROC curves for different \textit{initial dataset size}.}
\label{fig:initial_eer}
\end{figure}

We keep the local epochs $E$ fixed at 15 and take $\mathcal{P}_k = 320$ for $k = 1, 2$. We investigate the impact of different initial database sizes on the model's global performance. The initial model is trained as outlined in the section \ref{1-D CNN}. We set the scalar ratio as $r$, such that $|\mathcal{P}_{initial}| = r \sum |\mathcal{P}_{k}|$. We explore the search space for $\alpha$ which includes \{0\%, 5\%, 12.5\%, 25\%, 37.5\%, 50\%, 100\%\}. From Figure \ref{fig:initial_box} and Figure \ref{fig:initial_eer}, we can observe that larger initial database sizes result in better overall model performance.
Our experiments observed that even though the data used for initialization and the data from local databases do not follow the same distribution \textemdash as they belong to different users \textemdash larger initialization data leads to better model performance. \textbf{Our framework benefits from larger initialization data, resulting in improved transfer effects that guide the model in practical deployments}.

\subsubsection{\textbf{Scalability}}
The scalability of federated learning refers to its ability to adapt to large-scale data and multiple clients. In federated learning, many clients (such as mobile devices, sensors, cloud servers, etc.) train models locally and then share only the updated model weights rather than raw data. This decentralized model training approach provides federated learning with a degree of scalability, but it also comes with specific challenges and limitations. In this paper, we specifically emphasize scalability in terms of the number of clients.
\begin{table}[]
\centering
\begin{tabular}{c|cc}
\hline
Scenario             & EER     & Accuracy \\ \hline
centralized scenario & 3.33\%  & 96.25\%  \\
2\_agents\_FL        & 5.42\%  & 95.21\%  \\
5\_agents\_FL        & 5.83\%  & 94.17\%  \\
10\_agents\_FL       & 5.63\%  & 94.06\%  \\
20\_agents\_FL       & 10.63\% & 88.13\%  \\ \hline
\end{tabular}
\caption{Summary of scalability and comparison with the centralized scenario.}
\label{table:1}
\end{table}
\begin{figure}[h]
\centerline{\includegraphics[width=0.40\textwidth]{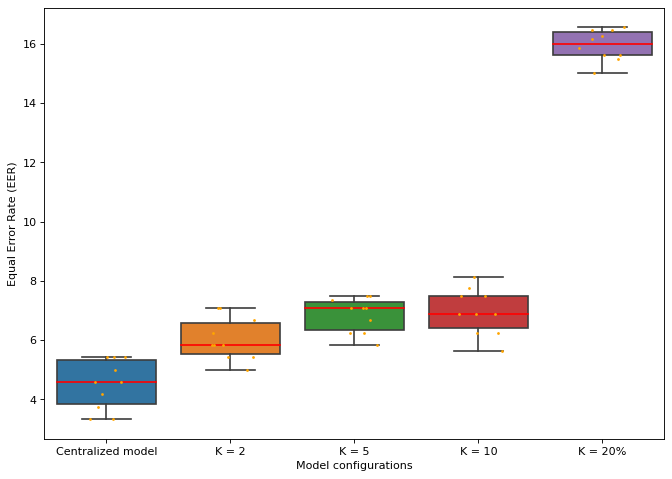}}
\caption{Boxplot for scalability.}
\label{fig:num_agents_box}
\end{figure}

\begin{figure}[h]
\centerline{\includegraphics[width=0.40\textwidth]{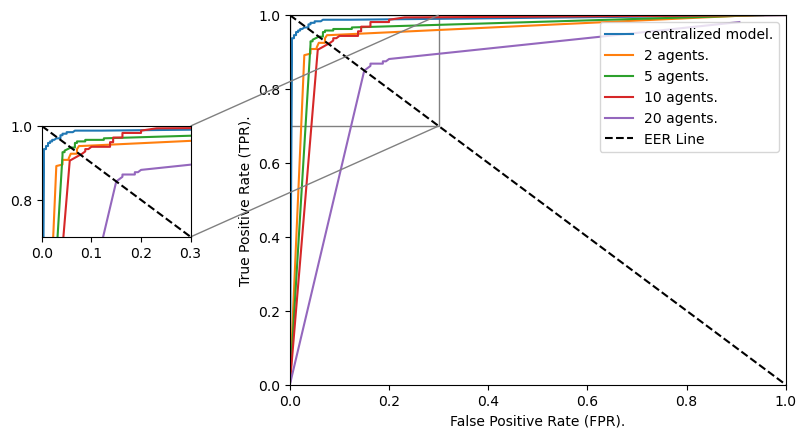}}
\caption{ROC curves for scalability.}
\label{fig:num_agents_EER}
\end{figure}
According to the results from the previous experiments, we set the number of \textit{local epochs} $E$ to 15 and the factor $r$ for the size of the data used for initialization to 100\%. Specifically, we use data from 40 out of the 80 users for initialization. In practical applications, we can consider the data used for initialization as contributed by volunteers or users who willingly participate in the local signature service. The remaining 40 users' signature data is evenly distributed among clients $k$. For example, when $k$ = 5, each agent receives data from 8 users. It is important to note that the test set is also derived from partial signature data of the 40 local users, maintaining a training-to-testing ratio of 8:2. Table \ref{table:1} summaries the scalability and comparison with the corresponding centralized scenario. \textbf{We can see from the Figure \ref{fig:num_agents_box} and \ref{fig:num_agents_EER} that our framework can achieve performance close to that of a centralized model with at least 10 agents, indicating that our framework can provide good scalability for at least 10 agents}.

\section{Conclusion}

In this paper, we have proposed a framework integrating a 1-D CNN into federated learning. Our framework prioritizes privacy preservation by enabling local training processes to maintain signature data within their respective personal environments. Besides, our framework achieves significant computational efficiency gains for central servers and local clients through a lightweight algorithm. Furthermore, our experiments highlight two key findings: 1) the remarkable effectiveness of the 1-D CNN in our framework for addressing online signature verification, and 2) the effectiveness of our adapted federated learning approach regarding lightweight local training, improved transfer effect, and high scalability.





\bibliographystyle{IEEEtran}
\bibliography{IEEEabrv,references}

\begin{thebibliography}{10}
\providecommand{\url}[1]{#1}
\csname url@samestyle\endcsname
\providecommand{\newblock}{\relax}
\providecommand{\bibinfo}[2]{#2}
\providecommand{\BIBentrySTDinterwordspacing}{\spaceskip=0pt\relax}
\providecommand{\BIBentryALTinterwordstretchfactor}{4}
\providecommand{\BIBentryALTinterwordspacing}{\spaceskip=\fontdimen2\font plus
\BIBentryALTinterwordstretchfactor\fontdimen3\font minus \fontdimen4\font\relax}
\providecommand{\BIBforeignlanguage}[2]{{%
\expandafter\ifx\csname l@#1\endcsname\relax
\typeout{** WARNING: IEEEtran.bst: No hyphenation pattern has been}%
\typeout{** loaded for the language `#1'. Using the pattern for}%
\typeout{** the default language instead.}%
\else
\language=\csname l@#1\endcsname
\fi
#2}}
\providecommand{\BIBdecl}{\relax}
\BIBdecl

\bibitem{2023_survey}
H.~Kaur and M.~Kumar, ``Signature identification and verification techniques: state-of-the-art work,'' \emph{Journal of Ambient Intelligence and Humanized Computing}, vol.~14, no.~2, pp. 1027--1045, 2023.

\bibitem{SSI}
Y.~Ding and H.~Sato, ``Self-sovereign identity as a service: architecture in practice,'' in \emph{2022 IEEE 46th Annual Computers, Software, and Applications Conference (COMPSAC)}.\hskip 1em plus 0.5em minus 0.4em\relax IEEE, 2022, pp. 1536--1543.

\bibitem{offline_1}
Y.~M. Al-Omari, S.~N. H.~S. Abdullah, and K.~Omar, ``State-of-the-art in offline signature verification system,'' in \emph{2011 International Conference on Pattern Analysis and Intelligence Robotics}, vol.~1.\hskip 1em plus 0.5em minus 0.4em\relax IEEE, 2011, pp. 59--64.

\bibitem{offline_2}
M.~M. Hameed, R.~Ahmad, M.~L.~M. Kiah, and G.~Murtaza, ``Machine learning-based offline signature verification systems: A systematic review,'' \emph{Signal Processing: Image Communication}, vol.~93, p. 116139, 2021.

\bibitem{online_1}
R.~Plamondon and S.~N. Srihari, ``Online and off-line handwriting recognition: a comprehensive survey,'' \emph{IEEE Transactions on pattern analysis and machine intelligence}, vol.~22, no.~1, pp. 63--84, 2000.

\bibitem{online_2}
D.~Impedovo and G.~Pirlo, ``Automatic signature verification: The state of the art,'' \emph{IEEE Transactions on Systems, Man, and Cybernetics, Part C (Applications and Reviews)}, vol.~38, no.~5, pp. 609--635, 2008.

\bibitem{online_3}
J.~Unar, W.~C. Seng, and A.~Abbasi, ``A review of biometric technology along with trends and prospects,'' \emph{Pattern recognition}, vol.~47, no.~8, pp. 2673--2688, 2014.

\bibitem{online_4}
A.~K. Jain, K.~Nandakumar, and A.~Ross, ``50 years of biometric research: Accomplishments, challenges, and opportunities,'' \emph{Pattern recognition letters}, vol.~79, pp. 80--105, 2016.

\bibitem{FL}
B.~McMahan, E.~Moore, D.~Ramage, S.~Hampson, and B.~A. y~Arcas, ``Communication-efficient learning of deep networks from decentralized data,'' in \emph{Artificial intelligence and statistics}.\hskip 1em plus 0.5em minus 0.4em\relax PMLR, 2017, pp. 1273--1282.

\bibitem{DTW}
J.~Picone, ``Fundamentals of speech recognition: A short course,'' \emph{Institute for Signal and Information Processing, Mississippi State University}, 1996.

\bibitem{DTW3}
L.~Nanni, E.~Maiorana, A.~Lumini, and P.~Campisi, ``Combining local, regional and global matchers for a template protected on-line signature verification system,'' \emph{Expert Systems with Applications}, vol.~37, no.~5, pp. 3676--3684, 2010.

\bibitem{DTW4}
A.~Sharma and S.~Sundaram, ``An enhanced contextual dtw based system for online signature verification using vector quantization,'' \emph{Pattern Recognition Letters}, vol.~84, pp. 22--28, 2016.

\bibitem{DTW1}
M.~Okawa, ``Time-series averaging and local stability-weighted dynamic time warping for online signature verification,'' \emph{Pattern Recognition}, vol. 112, p. 107699, 2021.

\bibitem{HMM2}
J.~Fierrez, J.~Ortega-Garcia, D.~Ramos, and J.~Gonzalez-Rodriguez, ``Hmm-based on-line signature verification: Feature extraction and signature modeling,'' \emph{Pattern recognition letters}, vol.~28, no.~16, pp. 2325--2334, 2007.

\bibitem{HMM3}
B.~L. Van, S.~Garcia-Salicetti, and B.~Dorizzi, ``On using the viterbi path along with hmm likelihood information for online signature verification,'' \emph{IEEE Transactions on Systems, Man, and Cybernetics, Part B (Cybernetics)}, vol.~37, no.~5, pp. 1237--1247, 2007.

\bibitem{RNN1}
S.~Lai, L.~Jin, and W.~Yang, ``Online signature verification using recurrent neural network and length-normalized path signature descriptor,'' in \emph{2017 14th IAPR international conference on document analysis and recognition (ICDAR)}, vol.~1.\hskip 1em plus 0.5em minus 0.4em\relax IEEE, 2017, pp. 400--405.

\bibitem{RNN2}
S.~Lai and L.~Jin, ``Recurrent adaptation networks for online signature verification,'' \emph{IEEE Transactions on information forensics and security}, vol.~14, no.~6, pp. 1624--1637, 2018.

\bibitem{RNN3}
G.~Li, L.~Zhang, and H.~Sato, ``In-air signature authentication using smartwatch motion sensors,'' in \emph{2021 IEEE 45th Annual Computers, Software, and Applications Conference (COMPSAC)}.\hskip 1em plus 0.5em minus 0.4em\relax IEEE, 2021, pp. 386--395.

\bibitem{normalize}
D.~Muramatsu and T.~Matsumoto, ``Effectiveness of pen pressure, azimuth, and altitude features for online signature verification,'' in \emph{Advances in Biometrics: International Conference, ICB 2007, Seoul, Korea, August 27-29, 2007. Proceedings}.\hskip 1em plus 0.5em minus 0.4em\relax Springer, 2007, pp. 503--512.

\bibitem{minibatch}
M.~Li, T.~Zhang, Y.~Chen, and A.~J. Smola, ``Efficient mini-batch training for stochastic optimization,'' in \emph{Proceedings of the 20th ACM SIGKDD international conference on Knowledge discovery and data mining}, 2014, pp. 661--670.

\bibitem{1d_CNN_survey}
S.~Kiranyaz, O.~Avci, O.~Abdeljaber, T.~Ince, M.~Gabbouj, and D.~J. Inman, ``1d convolutional neural networks and applications: A survey,'' \emph{Mechanical systems and signal processing}, vol. 151, p. 107398, 2021.

\bibitem{EEG1}
S.~Kiranyaz, T.~Ince, R.~Hamila, and M.~Gabbouj, ``Convolutional neural networks for patient-specific ecg classification,'' in \emph{2015 37th Annual International Conference of the IEEE Engineering in Medicine and Biology Society (EMBC)}.\hskip 1em plus 0.5em minus 0.4em\relax IEEE, 2015, pp. 2608--2611.

\bibitem{EEG2}
S.~Kiranyaz, T.~Ince, and M.~Gabbouj, ``Real-time patient-specific ecg classification by 1-d convolutional neural networks,'' \emph{IEEE Transactions on Biomedical Engineering}, vol.~63, no.~3, pp. 664--675, 2015.

\bibitem{DL}
I.~Goodfellow, Y.~Bengio, and A.~Courville, \emph{Deep learning}.\hskip 1em plus 0.5em minus 0.4em\relax MIT press, 2016.

\bibitem{BN}
S.~Ioffe and C.~Szegedy, ``Batch normalization: Accelerating deep network training by reducing internal covariate shift,'' in \emph{International conference on machine learning}.\hskip 1em plus 0.5em minus 0.4em\relax pmlr, 2015, pp. 448--456.

\bibitem{ReLU}
X.~Glorot, A.~Bordes, and Y.~Bengio, ``Deep sparse rectifier neural networks,'' in \emph{Proceedings of the fourteenth international conference on artificial intelligence and statistics}.\hskip 1em plus 0.5em minus 0.4em\relax JMLR Workshop and Conference Proceedings, 2011, pp. 315--323.

\bibitem{SGD}
H.~Robbins and S.~Monro, ``A stochastic approximation method,'' \emph{The annals of mathematical statistics}, pp. 400--407, 1951.

\bibitem{SGD2}
S.~Shalev-Shwartz, Y.~Singer, and N.~Srebro, ``Pegasos: Primal estimated sub-gradient solver for svm,'' in \emph{Proceedings of the 24th international conference on Machine learning}, 2007, pp. 807--814.

\bibitem{SVC-2004}
D.-Y. Yeung, H.~Chang, Y.~Xiong, S.~George, R.~Kashi, T.~Matsumoto, and G.~Rigoll, ``Svc2004: First international signature verification competition,'' in \emph{Biometric Authentication: First International Conference, ICBA 2004, Hong Kong, China, July 15-17, 2004. Proceedings}.\hskip 1em plus 0.5em minus 0.4em\relax Springer, 2004, pp. 16--22.

\bibitem{ROC}
T.~Fawcett, ``An introduction to roc analysis,'' \emph{Pattern recognition letters}, vol.~27, no.~8, pp. 861--874, 2006.

\bibitem{adam}
D.~P. Kingma and J.~Ba, ``Adam: A method for stochastic optimization,'' \emph{arXiv preprint arXiv:1412.6980}, 2014.

\end{thebibliography}


\end{document}